# Cavity Optofluidics: Raman Laser Made of a Fiber Coupled nano-Liter Droplet.

Shai Maayani[1, *], and Tal Carmon[2]

Research Laboratory of Electronics, Massachusetts Institute of Technology, Cambridge, MA 02139

Technion - Israel Institute of Technology, Faculty of Mechanical Engineering, Haifa 3200003, Israel

[*]corresponding author: maayani@mit.edu

## Abstract

**We fabricate a fiber coupled ultrahigh-Q resonator from a µdroplet per se, and experimentally measure stimulated Raman emission showing itself at a 160µW threshold. We observe Raman-laser lines that agree with their related calculated molecular-vibrations, as well as with a control-group experiment that uses a Raman-spectrometer. Yet, unlike spectrometers where emission is spontaneous and less directional, our droplet emitter is stimulated, single-mode-fiber coupled, resonantly enhanced with Q of 250 million, confined to a 23 $\mu m^3$ mode volume and has 7 orders of magnitude higher power.**

A droplet made of a simple oil, that was not designed to be an optically-active medium, is activated here as a fiber-coupled micro-laser emitter. Our optofluidic cavity constitute a bridge here, between the capillary properties of liquid interfaces, that facilitate constructing the atomically-smooth light-confining walls, while at the same time exploiting the optical-gain of this very-same oil, to permit laser emission. This optical amplification is achieved here via the Raman-gain mechanism that is common to all dielectrics. One can therefore say that the droplet actually construct itself to provide optical-gain and –feedback, needed for the laser. Furthermore and unlike cumbersome liquid-based dye lasers technology, our lasers footprint is only 56 microns.

From a broader view, the ability of liquids to dissolve, immerse, or contain analytes in a small volume of fluid was used for detecting nanoparticles[1,2] and molecules[3] as well as for sensing bio-analytes such as viruses[4] and proteins[5]. In most of such optofluidic sensors[3,6–8], cavities played a major rule in resonantly enhancing the interaction between light and matter while confining light to the analyte vicinity. Such cavities were activated as lasers emitters[9–12] to enhance detection. Generally speaking, the laser-enhanced sensitivity is achieved by relaying on Schawlow-Towens narrowing[13], that turns the laser linewidth narrower when compared with the cold-cavity linewidth. Additional enhancement of the sensitivity is achieved by a differential detection scheme where noise is cancel out.  This is done by relying on a split-mode, where the split is affected by the analyte while being relatively immune to noises such as the ones originating from thermal drifts[35]. In more details, two absorption deeps of a split mode[14,15] are measured[4–11,14–19] while considering only the drift of one resonance in respect to the other, which is proportional to signal. At the same time, noise-induced drifts, that are common to both resonances, are subtracted one from the other to cancel out.  One can therefore refer to such a sensor as benefiting from Schwalow-Townes narrowing in combination with a differential-detection capacity. Using Raman-laser[9–11] for such sensors[20] benefits also from the fact that there is no need to dope the dielectric for achieving laser emission. This allowed using a dopant-free ultrapure silica-glass, that has an ultrahigh Q[20]. It is therefore natural to ask why not to fabricate the entire resonator from liquid[9]. Mainly, because liquids have Raman gain, that makes them proper for dopant-free ultrahigh-Q Raman lasers[9]. Droplets were recently activated as ultrahigh-Q resonators while benefiting from being fiber-coupled and operating continuously in time [CW][24–29]. In addition, high index-contrast between the liquid and its air-cladding benefits total internal reflection near the air-liquid interface, with minimal radiation loss[31,32], while uniquely confining light to overlap almost entirely with the liquid core. A droplet laser can therefore benefit from several worlds, among them are small losses by radiation[31,32] and scattering[30] as well as tight confinement of light to almost perfectly overlap with the liquid.

Here we experimentally report on Raman-laser emission from liquid-walled optofluidic device in the form of a droplet resonator coupled to a tapered fiber.

*Fabrication* of the fiber coupled droplet resonator (Fig. 1) is done by dipping standard silica fiber with a fused ball on its end into a vessel containing silicone oil (polydimethylsiloxane) with a refractive index of 1.403.

*Activation* of the droplet as an optical resonator is done by evanescently coupling light to it using a tapered fiber[33,34]. The same tapered fiber is used, from both of its sides, to similarly couple the laser emission out. The scanning laser is a tunable 770-780 nm laser having a 300 kHz linewidth.

*Characterization* of the droplet optical quality is done by fast scanning the laser frequency through the droplet resonance-frequency and measuring the photon lifetime as indicated by its exponential decay (Fig. 2a). This measurement, generally referred to as ringdown, provides a quality factor of 250 million. Repeating this measurement at the frequency domain, by measuring the resonance linewidth (Fig. 2b), provides a quality factor of 160 million. In this measurement (Fig. 2b), the linewidth is probably broadened when measured in the frequency domain because of thermo-refractive resonance drift[35]. This is despite of operating at the under-coupled[34] regime to prevent thermal broadening[35]. As the ringdown measurements are by-definition taken at the temporal domain, they are proof against spectral broadening, including thermal ones. We therefore find the rigndown measurement (Fig. 2a), indicating a Q=250,000,000, as the one more reliable here.

*Experimentally* measuring the power and spectral characteristic of the Raman laser is done through both sides of the coupled fiber by connecting them to an optical spectrum analyzer (Advantest™ Q8384) or to a photodiode. A circulator is used to extract the backward Raman-lase from the fiber that brings the pump in, and direct it to the detector. At the same time, the forward Raman-laser is couple out, through the other side of the fiber, to a detector. To optimize the Raman-laser power we tune coupling by changing the gap between the fiber and droplet, using a nano-positioning system, until the Raman-laser power gets to a global maximum. The value of this coupling, where laser power is maximal, is generally referred to as optimal coupling[36]. As expected, optimal coupling is achieved at the slightly under-coupled regime[10,37].

*Raman-laser power* is measured here as a function of the pump input power. As one can see in figure 3, a knee shape reveals a threshold power of 160 µWatt for our Raman-laser. At this threshold, the Raman gain turns larger than loss. This knee separates between the spontaneous and the stimulated emission regions. As one can see in figure 5a, the power of our droplet Raman-laser is more than 7 orders of magnitude larger when compared to a Raman spectrometer where emission is spontaneous and less directional.

*Raman-laser spectrum* is measured using an optical spectrum analyzer to reveal several laser-lines (Fig. 4b) where each laser-line refers to a different vibration mode of the polydimethylsiloxane molecule (Fig. 4a). The spectrum is measured in backward direction while our droplet laser is optimally coupled. We compare the Raman spectrum to a calculation using a numerical-analysis software (SCM™ software, ADF module). As one can see in figure 4, (and watch in the supplementary movies), calculated molecular vibration modes falls within the bandwidth of the observed Raman-laser lines. The first observed Raman-laser line (S1) is related to a transverse vibration mode of the Oxygen atoms. The second observed Raman-laser line (S2) is related to a vibration mode of the hydrogen atoms. As expected, the higher-rate vibrational mode (S2) originates from motion of a lighter atom (Hydrogen), while the lower-rate optical phonon (S1) is associated with vibration of a heavier atom (Oxygen). The third observed mode (S3) is at a high rate that our computation resources could not cover. We estimate the third line (S3) to be involved in a mutual hydrogen molecule vibration, marked by a dash line in Fig. 4a.

We also compare the droplet-Raman-laser spectrum to a measurement done using a commercial Raman spectrometer - as a control-group experiment. As one can see in figure 5, our Raman lines (S1, S2 and an additional low frequency line) fall within the bandwidth of the lines measured by the commercial device. The high frequency of our third Raman laser-line (S3) falls out of the measuring range for the specific spectrometer-detector that we used.

To conclude, droplets represent one of the simplest and most-common forms of liquid in nature, industry and laboratory processes. We use droplets here as a bridge between the capillary properties of their interfaces, used here to construct the walls of our device, while at the same time using the liquid itself as an optical-gain medium. A micro-laser is then formed, when the droplet provide the structure as well as the optical-gain and –feedback.

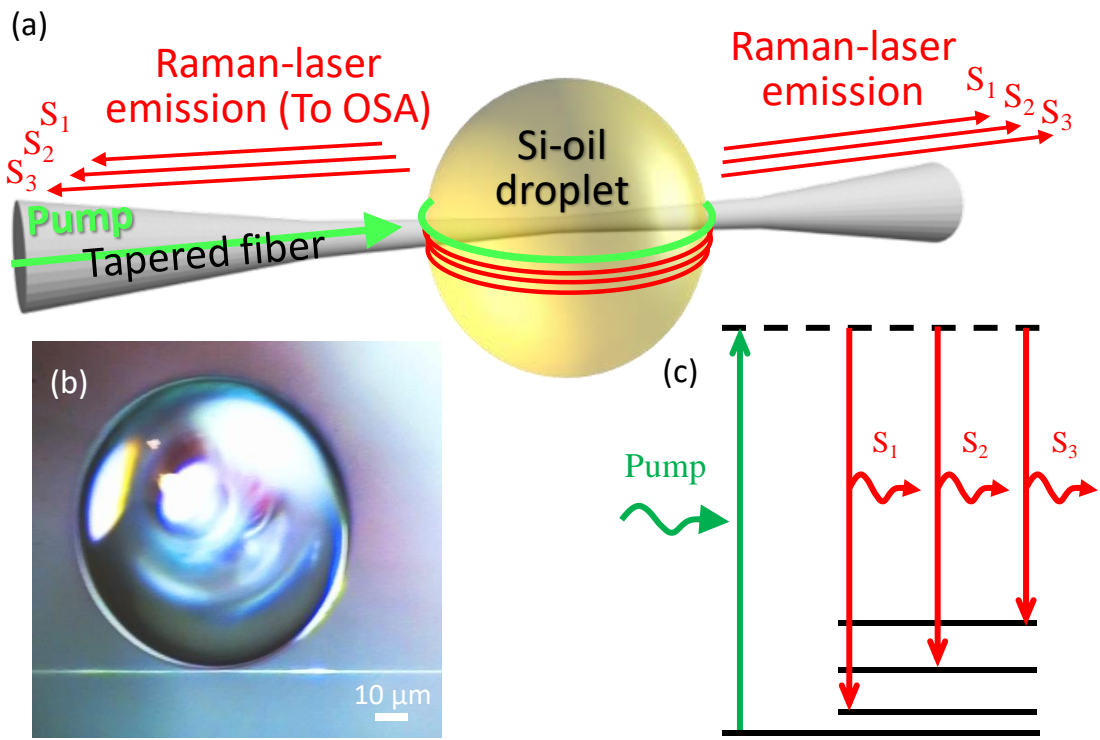

**Fig. 1. Experimental setup.** (a) Coherent emission of multiple Raman-laser lines from a liquid droplet resonator. The green arrow represents the pump, red arrows represent forward and backward stimulated Raman-emission. (b) Micrograph of our 78 µm diameter silicone oil droplet coupled to a tapered fiber (seen below). (c) Energy-level diagram illustrating three states involved in the Raman spectra.

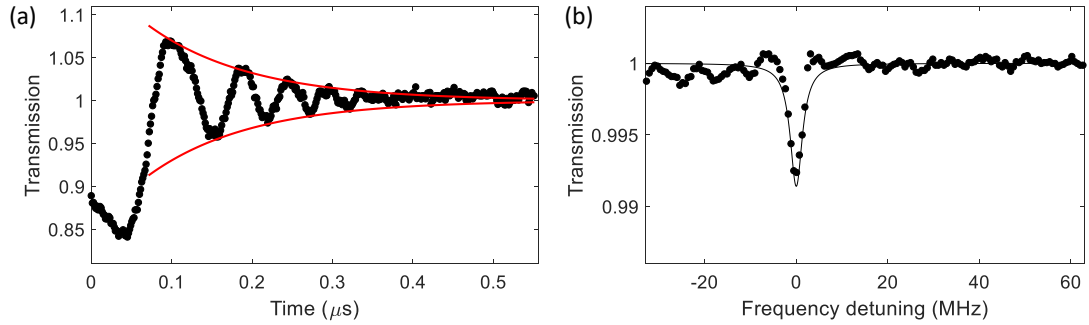

**Fig. 2. Measuring the optical quality factor.** (a) Scanning the pump laser through one of the resonances charges the resonator with light that decays later on. We fit an exponential decay (red) to provide the photon lifetime and measure a quality factor of $2.5 \cdot 10^8$. (b) Repeating this measurement in the frequency domain, while scanning relatively slow, reveals a Q of $1.6 \cdot 10^8$. We find measurement (a) more reliable since it is proof against broadening mechanisms.

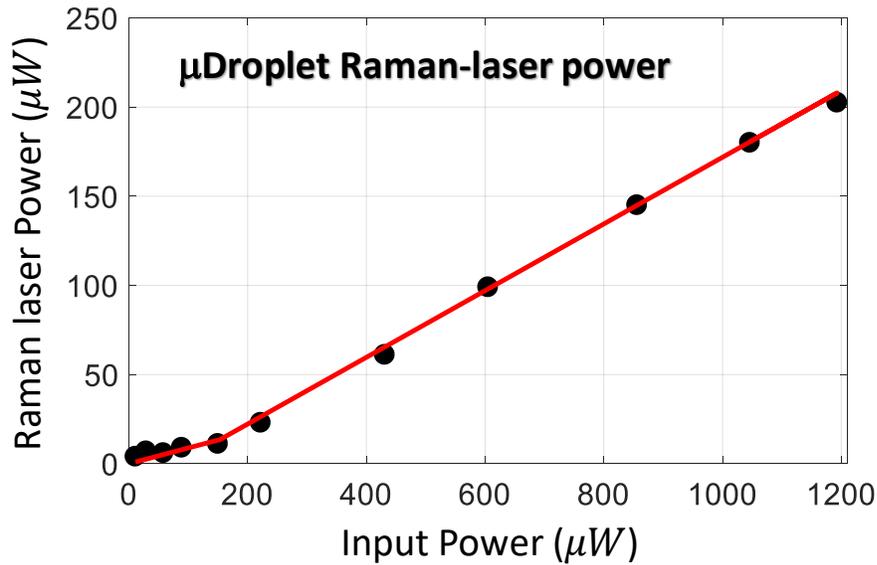

**Fig. 3. Experimentally measured threshold, power and efficiency for the microdroplet Raman laser.** Raman-laser power, out coupled via the fiber, as a function of the pump input power. The experimental data (circles) are fitted to the sum of two linear functions, one represents the spontaneous emission and the other represents the stimulated emission. A knee shape at 160 microwatt indicates the transition from spontaneous emission to stimulated emission, at input power generally referred as the lasing threshold. The slope efficiency here is 18%. R squared is 0.98 and 0.9995 for the spontaneous and stimulated fits. The size of the circles corresponds to the resolution of our measurement.

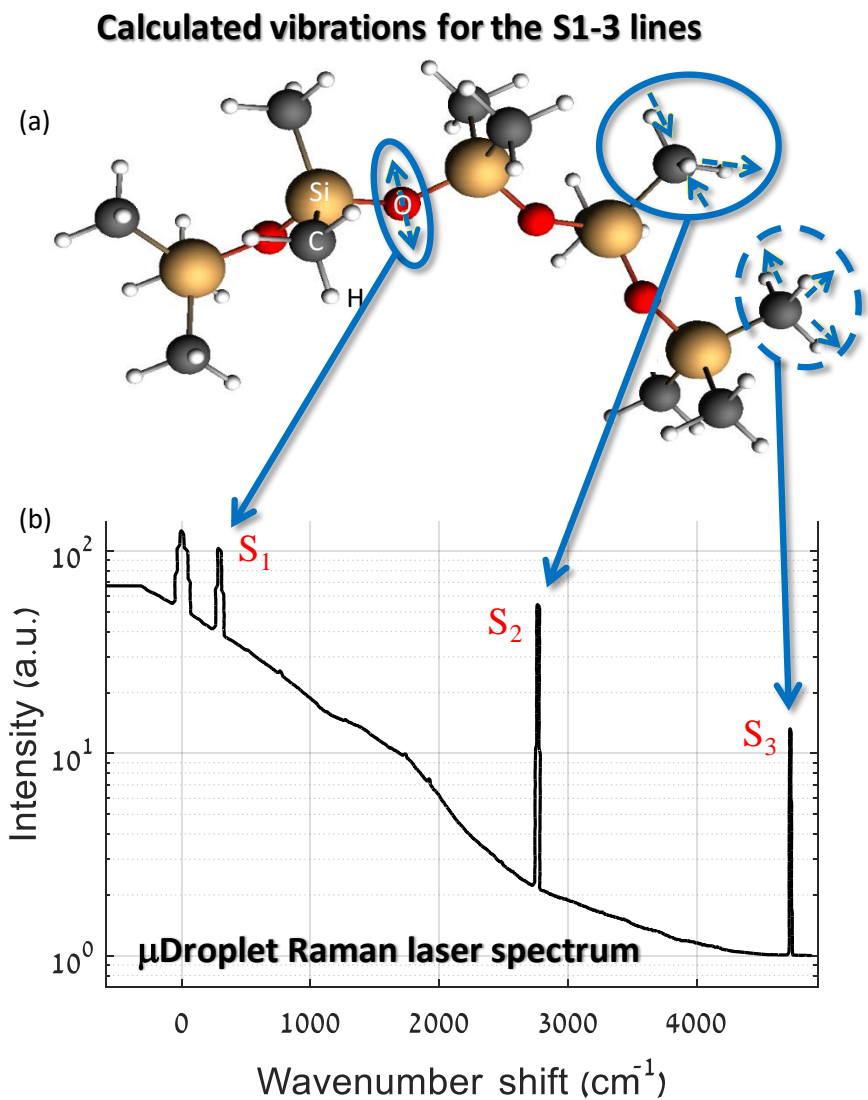

**Fig. 4. Experimentally measured Raman-laser lines and their corresponding calculated molecular vibrations.** (a) Numerical calculation of vibrational modes S1-2 of Polydimethylsiloxane with n=3 repeating monomer units. The vibrational modes were calculated by the ADF module of SCM™ software. Movies describing the dynamics involved in these S1-2 vibrations appear in the complementary part. The third vibrational

mode, S3, was estimated. (b) Raman spectrum of the droplet laser. The pump power and wavelength were 3mW and 778 nm. The optical spectrum analyzer was Advantest™ Q8384. The droplet is made from silicon-oil with viscosity of 60,000 mPa-s (85425 ALDRICH, chemical formula: $CH_3[Si(CH_3)_2O]_nSi(CH_3)_3$ ).

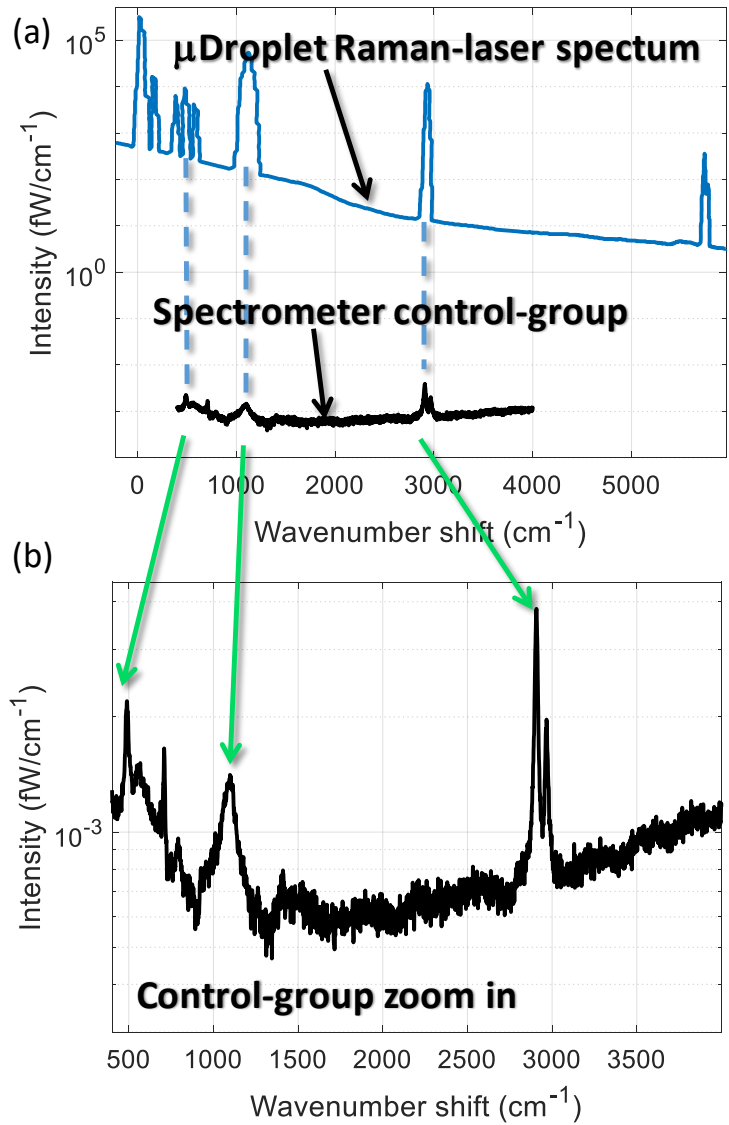

**Fig. 5.** Experimental results: (blue) Raman spectrum of 56-µm-diameter droplet resonator made from Silicone oil with viscosity of 1,000 mPa-s. (black) Control group: Raman spectrum obtained using a commercial Raman spectrometer (Horiba Jobin Yvon LabRAM HR Evolution®). The two spectra are provided together (a) for showing the higher power of the stimulated emission, then (b) contain zooming in to the control group experiment for providing its finer details.